\documentclass[aip,jcp,reprint,floatfix]{revtex4-2}

\usepackage[T1]{fontenc}
\usepackage[utf8]{inputenc}
\usepackage{newtxtext,newtxmath}

\usepackage{amsmath}
\usepackage{bm}
\usepackage{braket}
\usepackage[version=4]{mhchem}
\usepackage{siunitx}
\usepackage{dcolumn}
\usepackage{booktabs}
\usepackage{threeparttable}
\usepackage{xcolor}

\usepackage{graphicx}
\DeclareGraphicsExtensions{.pdf,.png,.jpg}

\providecommand{\selectlanguage}[1]{}

\usepackage{xurl}

\usepackage[
  colorlinks = true,
  linkcolor  = blue,
  citecolor  = blue,
  urlcolor   = blue,
  breaklinks = true
]{hyperref}

\begin{document}

\title{Direct Analytical Evaluation of Electron-Impact Excitation Cross Sections 
via Multiconfigurational Binary Encounter Approach: Applications to Benzene and Naphthalene}

\author{Kaoru Yamazaki}
\email{kaoru.yamazaki@aist.go.jp}
\affiliation{Materials DX Research Center, National Institute of Advanced Industrial Science and Technology (AIST), 
1-1-1 Umezono, Tsukuba, Ibaraki 305-8568, Japan}

\begin{abstract}
We present a multiconfigurational binary-encounter (MC-BE) framework
for direct analytical evaluation of electron-impact
electronic-excitation cross sections from \textit{ab initio}
excited-state data for dipole-allowed transitions. The method combines
the threshold-modified Mott--Massey (TMMM) approximation with
binary-encounter (BE/BE$f$) scaling. In this multiconfigurational
extension, the effective binding energy entering the BE/BE$f$
prefactor is evaluated from amplitude-weighted occupied-orbital
contributions computed by linear-response time-dependent density
functional theory (LR-TDDFT), without system-specific fitting
parameters. For benzene, MC-BE/TMMM cross sections for the dominant
$1\,{}^{1}\!E_{\mathrm{1u}}$ ($\pi\!\to\!\pi^{\ast}$) band agree
well with experiment over the 10--20\,eV range and, for this band and
range, show better agreement with experiment than the SMC/TCIS results
of Falkowski \textit{et al.} [J. Chem. Phys. \textbf{159}, 194301
(2023)]. For naphthalene, the calculated total excitation cross
section reproduces the onset and principal maximum of the gas-phase
apparent fluorescence cross section, used as an emission-based proxy
under dipole-dominated conditions, without empirical energy shifts or
intensity scaling. Analytic peak-position and peak-height expressions,
parameterized by $r=\langle B\rangle/\Delta E$, where $\Delta E$ is
the vertical excitation energy, indicate that typical valence
excitations peak at about $1.5$--$1.6\,\Delta E$ with substantial BE
attenuation, providing a practical diagnostic for relating measured
cross-section profiles to excitation energies. Although illustrated
with LR-TDDFT, the framework can be formally extended to other
wave-function-based excited-state theories, provided that compatible
amplitudes and well-defined orbital energies are available for the
chosen reference function. Together, the results support MC-BE/TMMM as
a practical, computationally inexpensive route for modeling
electron-impact excitation of polyatomic molecules for transitions
carrying finite oscillator strength.
\end{abstract}

\maketitle

\newcommand{\De}{\Delta E}
\newcommand{\Beff}{\langle B\rangle}
\newcommand{\sstar}{s^{\ast}}
\newcommand{\Tstar}{T^{\ast}}
\newcommand{\bfun}{\beta}

\section{Introduction}
\label{sec:introduction}

Electron-impact electronic-excitation cross sections are
fundamental quantities in gas-phase
scattering\cite{Tanaka_2016,Brunger_2002,Garcia-Abenza_2023}, plasma
kinetics\cite{Celiberto_2022}, and radiation
chemistry\cite{Sanche_2005}. 
Rigorous scattering approaches such as the
$R$-matrix method\cite{Tennyson_2010,Masin_2020}, the Schwinger multichannel
(SMC) method\cite{Lima_1990,daCosta_2015}, and the molecular convergent 
close-coupling method\cite{Scarlett_2022} can, in principle, provide accurate cross
sections across the full energy range; however, their computational
cost scales steeply with the number of electrons and target states,
limiting practical applications to atoms and small
molecules\cite{Tennyson_2010,Masin_2020,daCosta_2015,Scarlett_2022,Celiberto_2022}.
For medium-to-large polyatomic systems, computationally inexpensive
analytical formulations are therefore essential.

Analytical approaches rooted in the first Born picture---such as the
dipole-Born approximation (DBA)~\cite{Inokuti_1971}---yield compact
expressions that are accurate at high incident energies but require
threshold corrections and empirical adjustments to perform reliably
near the onset of excitation. The threshold-modified Mott--Massey
(TMMM) approximation~\cite{Celiberto_2013,Celiberto_2022} enforces the
correct threshold behavior and retains the proper high-energy
asymptote; nevertheless, it systematically overestimates integral
cross sections at low incident energies, motivating the introduction
of binary-encounter (BE) scaling and its oscillator-strength variant
(BE$f$)~\cite{Kim_2001,Tanaka_2016}. Yet direct application of these
models to \textit{ab initio} excited-state data has remained
impractical: standard binary-encounter (BE) formulations treat each
transition as originating from a single occupied orbital and therefore
cannot account for the multiconfigurational character of realistic
electronic excitations\cite{Kim_2001,Tanaka_2016}. In particular, no
principled procedure has existed to determine the effective binding
energy\cite{Kim_2001,Tanaka_2016}---the key quantity that governs the
BE/BE$f$ scaling---when multiple occupied orbitals contribute to a
given transition.

In this work, we remove this limitation by developing a
multiconfigurational binary-encounter (MC-BE) framework in which the
effective binding energy is evaluated from amplitude-weighted
occupied-orbital contributions obtained via linear-response
time-dependent density functional theory
(LR-TDDFT)\cite{Tsuneda_2014}. 
Together with excitation energies and oscillator strengths computed at
the same level of theory, this enables \textit{ab initio}
excited-state data to feed \emph{directly} into analytical
cross-section formulas for dipole-allowed transitions without
system-specific fitting. In the resulting MC-BE formulation, the TMMM
expression provides the threshold-corrected Born-type baseline, while
the BE/BE$f$ prefactor uses a multiconfigurational effective binding
energy evaluated from the excited-state amplitudes.

We validate the approach through benchmark calculations for benzene and
naphthalene, two prototypical aromatic hydrocarbons for which reliable
experimental electron energy-loss spectra (EELS) in the optical limit
and excitation cross-section data in the gas phase are available
\cite{Kato_2011,Falkowski_2023,Huebner_1972,McConkey_1992}. 
For benzene, the MC-BE/TMMM cross sections for the dominant
$1\,{}^{1}\!E_{\mathrm{1u}}$ ($\pi \to \pi^{\ast}$) band agree well with
experiment over the 10--20~eV range and, for this benchmark, show
better agreement with experiment than the Schwinger
multichannel/truncated configuration-interaction singles (SMC/TCIS)
results of Ref.~\cite{Falkowski_2023}.
For naphthalene, the calculated total
excitation cross section reproduces the onset and principal maximum of
the gas-phase fluorescence cross section\cite{McConkey_1992} without
empirical energy shifts or intensity scaling. 
For the specific benzene band investigated here, the approach is shown
to be weakly sensitive to the choice between the two long-range
corrected exchange--correlation functionals considered in
Appendix~\ref{app:xc_dependence}.
Although illustrated here with
LR-TDDFT, the methodology admits a formal extension to other
wave-function-based excited-state
theories\cite{Jensen_2017}, provided that compatible amplitudes and well-defined
orbital energies are available for the chosen reference function.
Moreover,
incorporation of Franck--Condon factors and Herzberg--Teller
vibronic coupling\cite{Santoro_2008,Barone_2009,Gozem_2022}
would account for vibronic effects beyond the present vertical
treatment and improve band-shape predictions, while spin--orbit
coupling\cite{Marian_2012,Marian_2021,Kotaru_2023,Kamiya_2018}
would allow transitions that acquire finite oscillator strength through
spin--orbit mixing to be treated.

\section{Theory}
\label{sec:theory}

This section presents the MC-BE formulation in two stages. We first
develop the cross-section expressions: the DBA\cite{Inokuti_1971} and
its threshold-corrected variant
(TMMM)\cite{Celiberto_2013,Celiberto_2022} provide the Born-type
baseline, and the BE/BE$f$ scaling supplies the threshold correction
used in the final cross section. The MC-BE extension enters through
the multiconfigurational evaluation of the effective binding energy
$\langle B_n\rangle$ from LR-TDDFT amplitudes
(Sec.~\ref{subsec:mcbe}). We then analyze the peak positions and peak
heights of both the TMMM and MC-BE cross sections as functions of the
dimensionless binding ratio
$r \equiv \langle B_n \rangle / \Delta E_n$
(Sec.~\ref{subsec:peak_analysis}), establishing diagnostic relations
that connect measurable cross-section features to excitation energies.
Full derivations are given in the Appendices.

Throughout this paper, we employ Hartree atomic units
($\hbar = m_\text{e} = a_0 = 1$,
1~Hartree~$\approx$~27.211~eV) unless otherwise specified.

\subsection{MC-BE cross-section formulation}
\label{subsec:mcbe}

Traditionally, analytical computation of electron-impact excitation
cross sections from excited-state calculations has been impractical
because standard BE models lack multiconfigurational
corrections. Here, we extend the BE approach to
incorporate LR-TDDFT amplitudes, enabling \emph{direct} evaluation of
cross sections from \textit{ab initio} excited-state data. The method
begins with the DBA\cite{Inokuti_1971}:
\begin{equation}
\sigma_n^{\mathrm{DBA}}(T) = \frac{2\pi g_n}{\Delta E_n\, T}\, f_n\,
\ln\!\left( \frac{k_f + k_i}{\left| k_f - k_i \right|} \right),
\label{eq:DBA}
\end{equation}
where $k_i$ and $k_f$ are the magnitudes of the initial and final
electron momenta, $T=k_i^2/2$ is the incident kinetic energy,
$\Delta E_n$ is the vertical excitation energy to 
the $n$-th electronically excited state
with degeneracy $g_n$, and $f_n$ is the oscillator strength.

To enforce the correct onset and retain analytic tractability, we use
the TMMM approximation\cite{Celiberto_2013,Celiberto_2022}
\begin{equation}
\sigma_n^{\mathrm{TMMM}}(T)
= \Theta(T-\Delta E_n)\,\frac{2\pi g_n}{\Delta E_n\, T}\, f_n\,
\ln\!\left( \frac{\sqrt{\Delta E_n}}
     {\sqrt{T} - \sqrt{T - \Delta E_n}} \right),
\label{eq:TMMM}
\end{equation}
where $\Theta(\cdot)$ is the Heaviside step function, included
here to define $\sigma_n^{\mathrm{TMMM}}(T)$ as identically
zero below threshold so that subsequent convolutions with
instrumental response functions used in
Sec.~\ref{subsec:naphthalene} are well-defined over the entire
energy axis.%
\footnote{The original TMMM
  formulation\cite{Celiberto_2013,Celiberto_2022} leaves
  $\Theta(T-\Delta E_n)$ implicit, since the argument of the
  logarithm is already real-valued only for
  $T \geq \Delta E_n$.}
Because the DBA and TMMM cross sections
(Eqs.~\eqref{eq:DBA} and \eqref{eq:TMMM}) are rooted in the
first Born (dipole) approximation, they are defined only for
optically allowed transitions with $f_n > 0$; for optically
forbidden transitions the formulas yield
$\sigma_n = 0$ by construction. Processes that proceed
through electron exchange or higher-multipole
interactions\cite{Celiberto_2022} lie outside the scope of the
present formulation.

At low incident energies, Eq.~\eqref{eq:TMMM} tends to overestimate
cross sections near threshold. To mitigate this, we apply
semiempirical BE and BE$f$ scalings~\cite{Kim_2001,Tanaka_2016}:
\begin{equation}
\sigma_n^{\text{BE/BE}f}(T) =
\frac{\Delta E_n}{\Delta E_n + \langle B_n \rangle + T}\;
\eta_n\;\sigma_n^{\mathrm{TMMM}}(T),
\label{eq:BE_BEf_scaling}
\end{equation}
where $\langle B_n \rangle$ is the effective binding energy of the
occupied-orbital manifold contributing to the transition, and $\eta_n$
scales the oscillator strength. For BE scaling, $\eta_n=1$, which is
the choice used for the results reported below; for BE$f$,
$\eta_n=f_n^{\text{ref}}/f_n^{\text{TMMM}}$, using a high-accuracy
reference $f_n^{\text{ref}}$ when available. Thus, the BE/BE$f$
prefactor provides the semiempirical scaling, whereas the present
MC-BE extension concerns the multiconfigurational evaluation of
$\langle B_n\rangle$ entering this prefactor. No system-specific
fitting parameters are introduced in this evaluation.
A key contribution of this work is a multiconfigurational
evaluation of $\langle B_n\rangle$ from LR-TDDFT excitation
amplitudes $X_{ia}^n$ and de-excitation amplitudes $Y_{ia}^n$,
where $i$ runs over occupied and $a$ over virtual
Kohn--Sham orbitals\cite{Klinkusch_2009,Coccia_2017,Wozniak_2024}:
\begin{equation}
\langle B_n\rangle
= -\frac{1}{C_n}\sum_{i}^{\text{occ}}\sum_{a}^{\text{vir}}
\left( |X_{ia}^n|^2 - |Y_{ia}^n|^2 \right)\,\epsilon_i,
\label{eq:Bn_LR_TDDFT}
\end{equation}
where $C_n$ is the normalization constant defined as\cite{Wozniak_2024}
\begin{equation}
C_n \equiv \sum_{i}^{\text{occ}}\sum_{a}^{\text{vir}}
\left( |X_{ia}^n|^2 - |Y_{ia}^n|^2 \right).
\label{eq:Bn_normalization_constant}
\end{equation}

The total excitation cross section is then
\begin{align}
\sigma_{\text{exc}}(T) &=
\sum_{n:\,\Delta E_n < \text{IP}}
\sigma_n^{\text{BE/BE}f}(T;\Delta E_n).
\label{eq:sigma_sum}
\end{align}
In practice, the sum in Eq.~\eqref{eq:sigma_sum} receives
contributions only from transitions carrying finite oscillator
strength ($f_n > 0$); optically forbidden states do not contribute
within the present vertical (Franck--Condon) framework. 
Incorporation of Franck--Condon factors, Herzberg--Teller vibronic
coupling\cite{Santoro_2008,Barone_2009,Gozem_2022}, and spin--orbit
coupling\cite{Marian_2012,Marian_2021,Kotaru_2023,Kamiya_2018} is
deferred to future work. These extensions would improve band-shape
predictions and allow transitions that acquire finite oscillator
strength through vibronic or spin--orbit coupling to be treated.
Equations~\eqref{eq:Bn_LR_TDDFT} and
\eqref{eq:Bn_normalization_constant} are specific to the LR-TDDFT
framework employed in this work. For general CI-type wave functions
that include double and higher excitations, analogous expressions
for the normalization constant $C_n$ (Eq.~\eqref{eq:Cn_CI}) and the
effective binding energy (Eq.~\eqref{eq:Bn_CI_general}) are derived
in Appendix~\ref{app:CI_Bn}\cite{Coccia_2017}. In the
single-excitation limit, these reduce to the CIS form
(Eq.~\eqref{eq:Bn_CIS}), which coincides with
Eq.~\eqref{eq:Bn_LR_TDDFT} under the Tamm--Dancoff approximation.
This generalization ensures that the MC-BE framework is applicable
to a broad range of excited-state theories, provided that well-defined
orbital energies are available for the chosen reference function.
Examples include equation-of-motion coupled-cluster (EOM-CC),
linear-response coupled-cluster (LR-CC), symmetry-adapted-cluster
configuration-interaction (SAC-CI), and algebraic diagrammatic
construction (ADC) methods\cite{Jensen_2017}.

\subsection{Peak analysis of TMMM and MC-BE cross sections}
\label{subsec:peak_analysis}

Having established the cross-section expressions, we now derive
analytic peak properties that serve as diagnostic tools for
interpreting experimental data.

For the TMMM cross section (Eq.~\eqref{eq:TMMM}), the peak position is
(see Appendix~\ref{sec:tmmm_peak} for derivation):
\begin{equation}
T^\star_{\mathrm{TMMM}} \approx 1.72349\,\Delta E_n,
\end{equation}
and the peak height scales as
\begin{equation}
\sigma^\star_{\mathrm{TMMM}} \approx
\frac{\mathcal{S}_{\mathrm{TMMM}}\,g_n f_n}{(\Delta E_n)^2},\qquad
\mathcal{S}_{\mathrm{TMMM}}\approx 2.8129.
\end{equation}

The BE/BE$f$ prefactor in Eq.~\eqref{eq:BE_BEf_scaling} shifts the
peak to lower energy and reduces its height relative to the TMMM
baseline. To quantify these effects, we define the dimensionless ratio
$r=\langle B_n\rangle/\Delta E_n$ and the peak multiplier
$s^*(r)=T^*/\Delta E_n$. 
The same-energy attenuation is
$R_{\mathrm{BE}}(r)=1/(1+r+s^*(r))$, and the peak-height ratio is
$\mathcal{R}_{\mathrm{peak}}(r)$ as defined in
Appendix~\ref{app:dim_peak}. Representative values are summarized in
Table~\ref{tab:rep_points}; the full $s^*(r)$ curve and the comparison
between $R_{\mathrm{BE}}(r)$ and $\mathcal{R}_{\mathrm{peak}}(r)$ are
shown in Figs.~\ref{fig:sstar_vs_r} and \ref{fig:R_both},
respectively.

Typical electronic excitations satisfy $r \sim 1$--$10$.
The peak of $\sigma_n^{\mathrm{BE/BE}f}(T)$ then appears at
$T \approx 1.5$--$1.6\,\Delta E_n$
(Table~\ref{tab:rep_points}), and its height ranges from
approximately $0.08$ to $0.28\,\sigma_{\mathrm{TMMM}}^{\ast}$
for $\eta_n = 1$. This relationship provides a practical means
of estimating $\Delta E_n$ from experimental
electronic-excitation cross sections.

\begin{table}[htbp]
\caption{Representative values of the peak multiplier $s^*(r)$,
  same-energy attenuation $R_{\mathrm{BE}}(r)$, and peak-height
  ratio $\mathcal{R}_{\mathrm{peak}}(r)$ for selected binding
  ratios $r$ (four significant figures, $\eta_n = 1$).}
\label{tab:rep_points}
\centering
\begin{tabular}{lccc}
\toprule
$r$ & $s^*(r)$ & $R_{\mathrm{BE}}(r)$ &
  $\mathcal{R}_{\mathrm{peak}}(r)$\\
\midrule
$0$        & $1.3955$ & $0.4174$ & $0.3965$ \\
$1$        & $1.4528$ & $0.2896$ & $0.2806$ \\
$10$       & $1.6128$ & $0.0793$ & $0.0790$ \\
$\infty$   & $1.7235$ & ---      & --- \\
\bottomrule
\end{tabular}

\vspace{3pt}
\footnotesize
For $r\to\infty$, both ratios vanish as
$R_{\mathrm{BE}}(r)\sim \big(1+r+s_\infty\big)^{-1}$
and $\mathcal{R}_{\mathrm{peak}}(r)\sim \big(1+r+s_\infty\big)^{-1}$
with $s_\infty=1.7235$.
\end{table}

The peak multiplier $s^{\ast}(r)$ is strictly increasing with the
binding ratio $r$. In the small-$r$ regime it grows linearly with
slope $0.07147$, i.e.,
$s^{\ast}(r) \approx s^{\ast}(0) + 0.07147\,r$. In the opposite limit
it approaches $s_{\infty}=1.7235$ from below with a leading
inverse-$r$ behavior,
$s^{\ast}(r) \approx s_{\infty} - 1.7448/r$. Derivations of these
coefficients are provided in Appendix~\ref{app:dim_peak}.

\section{Computational Details}
\label{sec:comp_details} 

Equilibrium geometries of benzene and naphthalene were optimized at the
B3LYP/def2-TZVP level of density functional
theory\cite{Becke_1993,Weigend_2005}. Normal-mode analyses were
performed at the same level to confirm convergence to minima.

Singlet excited-state wave functions, excitation energies, and
oscillator strengths for 3000 low-lying states were computed by
LR-TDDFT using the Casida formalism\cite{Tsuneda_2014,Jensen_2017}. We
employed the long-range corrected hybrid functional
$\omega$B97X-D\cite{Chai_2007} to describe both valence and Rydberg
excitations and to provide reliable orbital energies and ionization
potentials (IPs), which are essential for evaluating
$\langle B_n\rangle$ via Eq.~\eqref{eq:Bn_LR_TDDFT}, as summarized in
Table~\ref{tab:ionization_potentials} in the Appendix~\ref{app:ip_dft_vs_exp}. To compute
$\langle B_n\rangle$, we included excitation configurations satisfying
$10^{-4}\le |X_{ia}^n| \le 1$ and de-excitation configurations
satisfying $10^{-4}\le |Y_{ia}^n| \le 1$.

For C, we used the Sapporo-2012-TZP\cite{Noro_2012} basis
set augmented with diffuse functions $(1s,1p,1d,1f)$ and further
augmented by the Dunning--Hay double-Rydberg set
$(2s,2p,2d)$\cite{Dunning_1977} to balance valence and Rydberg
character. For H, Sapporo-2012-DZP\cite{Noro_2012} was employed.

The IP was evaluated by the $\Delta$SCF method at the same level of
theory as the LR-TDDFT calculations.

All quantum-chemistry calculations were carried out with
\textit{Gaussian~16} Rev.~C.02\cite{g16}. Basis sets were obtained
from the \textit{Basis Set
Exchange}\cite{Pritchard_2019,Feller_1996,Schuchardt_2007}.

\section{Results and Discussion}
\label{sec:results}

To validate the MC-BE/TMMM framework developed in
Sec.~\ref{sec:theory}, we apply it to the electron-impact excitation
of two prototypical aromatic hydrocarbons: benzene and naphthalene.
These molecules were chosen because (i) reliable gas-phase
experimental data are available---an electron energy-loss spectrum
(EELS)\cite{Kato_2011} and an integral excitation cross
section\cite{Falkowski_2023} for benzene, and an
EELS\cite{Huebner_1972} and a fluorescence cross
section\cite{McConkey_1992} for naphthalene---and (ii) progression
from a monocyclic to a fused polycyclic system probes the ability of
the method to handle an increasingly dense manifold of low-lying
excited states. For each molecule, we first assess the quality of the
TD-$\omega$B97X-D excitation energies and oscillator strengths by
comparing the simulated EELS with experiment, thereby confirming that
the \textit{ab initio} input to the MC-BE formulas is physically
reasonable. We then evaluate the MC-BE/TMMM cross sections against the
available experimental cross-section data and identify the dominant
contributing excited states through state-resolved analysis.

\begin{figure*}[t]
  \centering
  \includegraphics[width=\textwidth]{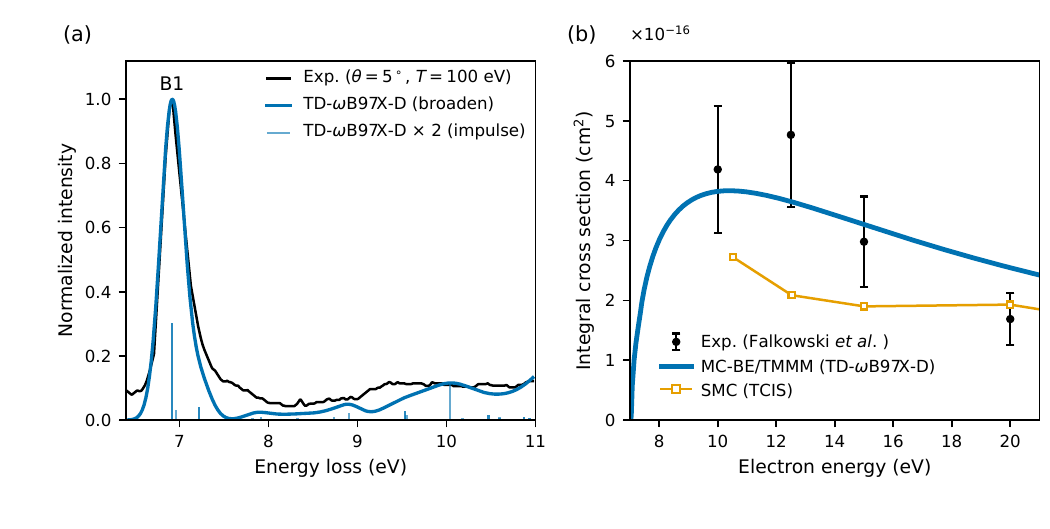}
  \caption{(Color online)
(a) Electron energy-loss spectrum (EELS) of benzene at 100~eV incident
energy and $5^\circ$ scattering angle (black), normalized in the
6.4--11.0~eV loss range\cite{Kato_2011}. The simulated valence
excitation spectrum in the optical limit (blue) is based on 181
optically allowed singlet states from TD-$\omega$B97X-D
($<15$\,eV), broadened by Gaussians (FWHM = 0.3\,eV below and
0.6\,eV above the ionization potential, 9.27\,eV). Vertical sticks
show intensities scaled by 2 (impulse approximation). The calculated
spectrum is shifted by $-0.13$\,eV only for visual alignment of the
dominant B1 feature in panel (a) with experiment.
(b) Integral cross section for electron-impact excitation in the B1
band of benzene in panel (a) for $T = 7$--21~eV. Experimental data
(black circles, Ref.~\cite{Falkowski_2023}) are compared with the
present MC-BE/TMMM results (blue) and the SMC/TCIS results reported
by Falkowski \textit{et al.}\cite{Falkowski_2023} (orange squares).}
  \label{fig:benzene_eels_and_cross_section}
\end{figure*}

\subsection{Benzene}

The electron-impact excitation spectrum of benzene provides a
well-defined benchmark for the present MC-BE analysis of polyatomic
molecules. In this subsection, we compare the experimental
EELS\cite{Kato_2011} and the gas-phase B1-band excitation
cross section\cite{Falkowski_2023} with calculated excitation
properties, and demonstrate the capability of the MC-BE approximation
to reproduce the excitation cross section of dipole-allowed
transitions.
Because the present EELS comparison is taken at $\theta=5^{\circ}$ and
$T=100$~eV, the optical-limit approximation at small momentum transfer
$q$ is a reasonable working assumption, rather than as a full
finite-$q$ EELS treatment.

\subsubsection{Overview of benzene excitation features in EELS}

Figure~\ref{fig:benzene_eels_and_cross_section}(a) displays the
experimental EELS spectrum of benzene recorded at an incident electron
energy of $T = 100$~eV and a scattering angle of
$\theta = 5^{\circ}$.\cite{Kato_2011}
The spectrum exhibits a prominent dipole-allowed feature, hereafter
labeled B1 ($\sim$6.9~eV), which encompasses contributions from one or
more singlet excited states.

To facilitate comparison, a simulated spectrum was constructed
from the TD-$\omega$B97X-D vertical excitation energies
(Sec.~\ref{sec:comp_details}) by convolving each transition
with a Gaussian line shape of FWHM $= 0.30$ and $0.60$~eV
for singlet excited states below and above the $\Delta$SCF
ionization potential (9.27~eV), respectively.
A uniform energy shift of $-0.13$~eV was applied to align the
calculated envelope with the experimental spectrum. The
resulting stick spectrum and the broadened profile are
superimposed on the experimental EELS trace in
Fig.~\ref{fig:benzene_eels_and_cross_section}(a).

On the basis of both the present TD-$\omega$B97X-D results and
previous calculations at the CASPT2\cite{Lorentzon_1995},
EOM-CC3\cite{Falkowski_2021}, and FC/SSE(SAC--CI,
ESF)\cite{Nakatsuji_2026} levels of theory
(Table~\ref{tab:benzene_tddft_eomcc3_exp}), the B1 band is attributed
to the $1^1\!E_\mathrm{1u}$, $2^1\!E_\mathrm{1u}$, and
$1^1\!A_\mathrm{2u}$ states. The four theoretical methods agree with
one another and with experiment to within $\sim$0.2~eV for the band
positions (Table~\ref{tab:benzene_tddft_eomcc3_exp}). The vertical
excitation energies and oscillator strengths therefore provide
physically reasonable input for the subsequent electron-impact
excitation cross-section calculations within the MC-BE formalism. The
ionization potential calculated from the orbital energy of the HOMO
(9.07~eV) quantitatively agrees with that from $\Delta$SCF
(9.27~eV) and with experiment
(9.24~eV\cite{Asbrink_1970}) as summarized in
Table~\ref{tab:ionization_potentials} in the Appendix~\ref{app:ip_dft_vs_exp}. 
This indicates that the calculated
orbital energies are reasonable input for evaluating the effective
binding energy $\langle B_n \rangle$ in the MC-BE formalism.

\begin{table*}[!t]
  \caption{Comparison of vertical excitation energies ($\Delta E_n$, in eV) and oscillator strengths ($f_n$) for the three lowest optically allowed singlet excited states of benzene, obtained using TD-$\omega$B97X-D, CASPT2\cite{Lorentzon_1995}, EOM-CC3\cite{Falkowski_2021}, FC/SSE(SAC--CI, ESF)\cite{Nakatsuji_2026}, and experimental data\cite{Kato_2011,Lorentzon_1995,Kitao_1987}.}
  \label{tab:benzene_tddft_eomcc3_exp}
  \centering
  \small
  \setlength{\tabcolsep}{3.0pt}
  \begin{threeparttable}
  \begin{tabular}{
    l
    S[table-format=2.3] S[table-format=1.3,round-precision=3]
    S[table-format=2.3] S[table-format=1.3,round-precision=3]
    S[table-format=2.3]
    S[table-format=2.3] S[table-format=1.3,round-precision=3]
    S[table-format=2.3] S[table-format=1.3]
  }
    \toprule
    \multicolumn{1}{c}{State} &
    \multicolumn{2}{c}{TD-$\omega$B97X-D\textsuperscript{a}} &
    \multicolumn{2}{c}{CASPT2\textsuperscript{b}} &
    \multicolumn{1}{c}{EOM-CC3\textsuperscript{c}} &
    \multicolumn{2}{c}{FC/SSE(SAC--CI, ESF)\textsuperscript{d}} &
    \multicolumn{2}{c}{Experiment\textsuperscript{e}} \\
    \cmidrule(r){2-3}
    \cmidrule(r){4-5}
    \cmidrule(r){6-6}
    \cmidrule(r){7-8}
    \cmidrule(r){9-10}
    {} &
    {$\Delta E_n$ (eV)} & {$f_n$} &
    {$\Delta E_n$ (eV)} & {$f_n$} &
    {$\Delta E_n$ (eV)} &
    {$\Delta E_n$ (eV)} & {$f_n$} &
    {$\Delta E_n$ (eV)} & {$f_n$} \\
    \midrule

    $1^1\!A_\mathrm{2u}$ ($\pi \to 3p_\sigma$) &
      7.09 & 0.056 &
      6.86 & 0.052 &
      6.91 &
      6.95 & 0.087 &
      6.93 & \multicolumn{1}{c}{---} \\

    $1^1\!E_\mathrm{1u}$ ($\pi \to \pi^\ast$) &
      7.05 & 1.085 &
      7.03 & 0.820 &
      7.01 &
      7.09 & 0.468 &
      6.94 & \multicolumn{1}{c}{0.824--0.953} \\

    $2^1\!E_\mathrm{1u}$ ($\pi \to 3p_\pi$) &
      7.35 & 0.145 &
      7.16 & 0.058 &
      7.26 &
      7.39 & 0.514 &
      7.41 & \multicolumn{1}{c}{---} \\

    \bottomrule
  \end{tabular}

  \vspace{4pt}
  \begin{tablenotes}
    \raggedright
    \footnotesize
    \item \textsuperscript{a} This work.
    \item \textsuperscript{b} Lorentzon \textit{et al.}\cite{Lorentzon_1995}: [6e,13o]-CASPT2/ANO-S + single-center Rydberg $(8s,8p,8d) \rightarrow [1s,1p,1d]$ at the center of mass.
    \item \textsuperscript{c} Falkowski \textit{et al.}\cite{Falkowski_2021}: EOM-CC3/C: d-aug-cc-pVDZ, H: aug-cc-pVDZ.
    \item \textsuperscript{d} Nakatsuji \cite{Nakatsuji_2026}: FC/SSE(SAC--CI, ESF) theory at the L6 level.
    \item \textsuperscript{e} Refs.~\cite{Kato_2011,Lorentzon_1995,Kitao_1987}, and references therein. A relative uncertainty of $\sim$20\% is expected for experimental oscillator strengths~\cite{Kato_2011}.
  \end{tablenotes}
  \end{threeparttable}
\end{table*}

\subsubsection{Integral excitation cross section}

The gas-phase integral excitation cross section for the B1 band serves
as an experimental reference for validating the calculated cross
sections. For a typical electronic excitation with $r \sim 1$--$10$,
the cross-section maximum for a dipole-allowed transition is
expected to appear at an incident energy of approximately
$1.5$--$1.6\,\Delta E$
(Table~\ref{tab:rep_points}). On this basis, the main feature of the
experimentally observed cross section at 12.5~eV is assigned to the
dipole-allowed $1^1\!E_\mathrm{1u}$, $2^1\!E_\mathrm{1u}$, and
$1^1\!A_\mathrm{2u}$ states.

The calculated total excitation cross section is compared with the
experimental excitation cross section in
Fig.~\ref{fig:benzene_eels_and_cross_section}(b). The overall
agreement is satisfactory: the theoretical profile reproduces the
behavior in the range $T = 10$--20~eV. 
Furthermore, for this B1-band benchmark over the 10--20~eV range, the
MC-BE/TMMM results show better agreement with experiment than the
SMC/TCIS results of Ref.~\cite{Falkowski_2023}.
For the benzene B1-band benchmark examined here, the calculated
cross section is weakly sensitive to the choice between the two
long-range corrected exchange--correlation functionals considered in
Appendix~\ref{app:xc_dependence}
(Fig.~\ref{fig:xc_functional_dependence} and Table~\ref{tab:benzene_peaks_xc_dependence}).

\begin{figure}[t]
  \centering
  \includegraphics[width=\linewidth]{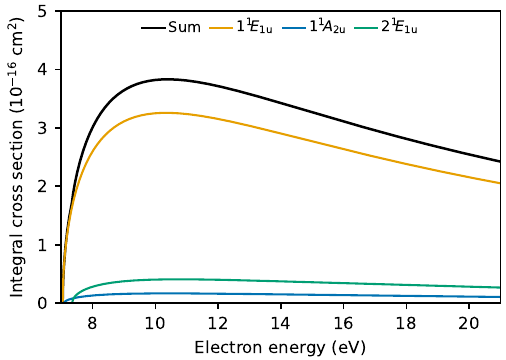}
  \caption{(Color online)
  The MC-BE/TMMM calculated electron-impact excitation cross sections of benzene
  using TD-$\omega$B97X-D target states as a function of incident
  electron energy $T$. Black line: sum of the
  $2^{1}E_{\mathrm{1u}}$, $1^{1}E_{\mathrm{1u}}$, and
  $1^{1}A_{\mathrm{2u}}$ states; orange line:
  $2^{1}E_{\mathrm{1u}}$ state; blue line:
  $1^{1}E_{\mathrm{1u}}$ state; green line:
  $1^{1}A_{\mathrm{2u}}$ state.}
  \label{fig:benzene_ics_mcbe_smc}
\end{figure}

Analysis of the state-resolved cross sections
(Fig.~\ref{fig:benzene_ics_mcbe_smc}) confirms that the dipole-allowed
$1\,^{1}\!E_{\mathrm{1u}}$ ($\pi \to \pi^{\ast}$) transition is the
dominant contributor, accounting for 85\% of the total cross
section at its peak position ($T = 10.39$~eV). 
Because the $1\,{}^{1}\!E_{\mathrm{1u}}$ and $2\,{}^{1}\!E_{\mathrm{1u}}$ states form 
a closely spaced valence--Rydberg pair, this state-resolved percentage should be regarded as functional-specific; 
the total B1-band excitation cross section is the more robust
quantity, as illustrated by the functional comparison in
Fig.~\ref{fig:xc_functional_dependence} and
Table~\ref{tab:benzene_peaks_xc_dependence}. The threshold of this
state is located at $T = 7.05$~eV.

\subsection{Naphthalene}
\label{subsec:naphthalene}

The electron-impact excitation spectrum of naphthalene provides an
additional benchmark for extending the present MC-BE analysis from
monocyclic to fused polycyclic aromatic systems. Compared with
benzene, naphthalene possesses a considerably denser manifold of
low-lying electronic states\cite{Kannar_2014}, rendering the
assignment of individual spectral features more challenging. In this
subsection, we compare the experimental EELS\cite{Huebner_1972} and
the gas-phase fluorescence cross
section\cite{McConkey_1992}---which is approximately proportional to
the total excitation cross section\cite{Kazakov_2000}---with
calculated excitation properties, and demonstrate the capability of
the MC-BE approximation to identify the dominant contributing excited
states underlying the observed spectral features.

\begin{figure*}[t]
  \centering
  \includegraphics[width=\textwidth]{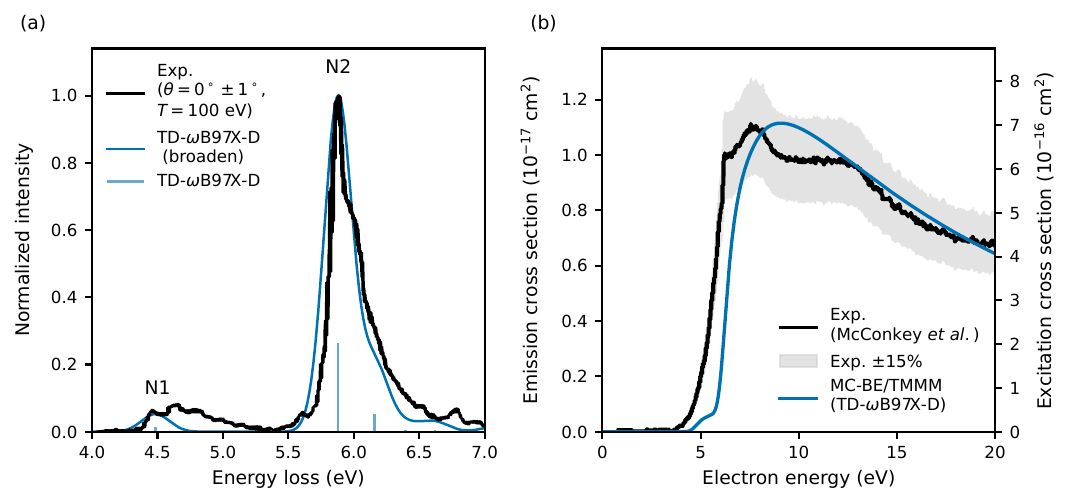}
  \caption{
Experimental and calculated electron-impact excitation spectra of
naphthalene. (a) EELS spectrum at $T = 100$~eV and
$\theta = 0^\circ \pm 1^\circ$\cite{Huebner_1972} compared with
TD-$\omega$B97X-D excitation energies. The black curve is the
experimental spectrum, the pale-blue vertical lines are the
calculated impulse spectrum, and the blue curve is the
area-normalized Gaussian-broadened theoretical spectrum. The
theoretical EELS profile was shifted and intensity-scaled only for
visual alignment of the dominant N2 feature in panel (a); N1 and N2
denote the main spectral features. (b) Experimental apparent emission
cross section\cite{McConkey_1992} for the naphthalene fluorescence band
compared with the MC-BE/TMMM calculated total excitation cross
section using TD-$\omega$B97X-D target states. The black curve and
gray band denote the experimental central values and
$\pm 15\%$ uncertainty\cite{McConkey_1992}, respectively. The
calculated cross section was convoluted with an area-normalized
Gaussian of FWHM $= 0.50$~eV, corresponding to the reported
electron-beam energy spread\cite{McConkey_1992}, and is shown without
energy shifting or intensity scaling; the right ordinate gives the
unscaled calculated values after broadening. The fluorescence cross
section is used as an indirect reference for the dipole-dominated
total excitation profile.}
  \label{fig:naphthalene_eels_cross_section}
\end{figure*}

\subsubsection{Overview of naphthalene excitation features in EELS}

Figure~\ref{fig:naphthalene_eels_cross_section}(a) displays the
experimental EELS spectrum of naphthalene recorded at an incident
electron energy of $T = 100$~eV and a scattering angle of
$\theta = 0^{\circ} \pm 1^{\circ}$\cite{Huebner_1972}. The spectrum
exhibits two prominent loss-energy regions, hereafter labeled N1
($\sim$4.5~eV) and N2 ($\sim$5.9~eV), each of which encompasses
contributions from one or more singlet excited states.

To facilitate comparison, a simulated spectrum was constructed
from the TD-$\omega$B97X-D vertical excitation energies
(Sec.~\ref{sec:comp_details}) by convolving each transition
with a Gaussian line shape of FWHM $= 0.25$~eV and applying
a uniform energy shift of $-0.18$~eV to align the calculated
envelope with the experimental spectrum.
The resulting stick spectrum and the broadened profile are
superimposed on the experimental EELS trace in
Fig.~\ref{fig:naphthalene_eels_cross_section}(a).

On the basis of both the present TD-$\omega$B97X-D results and
previous calculations at the CASPT2\cite{Rubio_1994} and LR-CC levels
of theory\cite{Falden_2009,Kannar_2014}
(Table~\ref{tab:naphthalene_tddft_eomcc3_exp}), the two loss-energy
regions are assigned as follows. The N1 band centered at
$\sim$4.5~eV is attributed to the $1\,^{1}\!B_\mathrm{2u}$
($\pi \to \pi^{\ast}$) state, whereas the N2 band near 5.9~eV arises
from the closely spaced $2\,^{1}\!B_\mathrm{3u}$ ($\pi \to \pi^{\ast}$) and
$2\,^{1}\!B_\mathrm{2u}$ ($\pi \to \pi^{\ast}$) states. The four theoretical
methods agree with one another and with experiment to within
$\sim$0.3~eV for the band positions
(Table~\ref{tab:naphthalene_tddft_eomcc3_exp}).

The principal discrepancy is that the simulated N1 band is
appreciably narrower than the experimental profile. This narrowing is
a direct consequence of the vertical (Franck--Condon) approximation
adopted here, which neglects the vibronic envelope arising from the
Franck--Condon overlap and Herzberg--Teller vibronic
coupling
\cite{Atkins_2011,Santoro_2008,Barone_2009,Gozem_2022} between the
ground- and excited-state potential-energy surfaces. Despite this
limitation, the vertical excitation energies and oscillator strengths
provide physically reasonable input for the subsequent
electron-impact excitation cross-section calculations within the MC-BE
formalism. The ionization potential calculated from the orbital energy
of the HOMO (7.98~eV) quantitatively agrees with that from
$\Delta$SCF (8.04~eV) and with experiment
(8.00~eV\cite{Yamauchi_1998}) as summarized in
Table~\ref{tab:ionization_potentials} in the Appendix~\ref{app:ip_dft_vs_exp}. 
This indicates that the calculated
orbital energies are reasonable input for evaluating the effective
binding energy $\langle B_n \rangle$ in the MC-BE formalism.

\begin{table*}[t]
  \caption{Comparison of vertical excitation energies ($\Delta E_n$, in eV) and oscillator strengths ($f_n$)
  for the three lowest optically allowed singlet excited states of naphthalene, obtained using TD-$\omega$B97X-D,
  CASPT2\cite{Rubio_1994}, LR-CCSDR(3)\cite{Falden_2009}, LR-CC3\cite{Kannar_2014}, and experimental data\cite{Huebner_1972}.}
  \label{tab:naphthalene_tddft_eomcc3_exp}
  \centering
  \small
  \setlength{\tabcolsep}{3.0pt}
  \begin{threeparttable}
  \begin{tabular}{
    l
    S[table-format=2.3] S[table-format=1.3,round-precision=3]
    S[table-format=2.3] S[table-format=1.3,round-precision=3]
    S[table-format=2.3]
    S[table-format=2.3] S[table-format=1.3,round-precision=3]
    S[table-format=2.3] S[table-format=1.3]
  }
    \toprule
    \multicolumn{1}{c}{State} &
    \multicolumn{2}{c}{TD-$\omega$B97X-D\textsuperscript{a}} &
    \multicolumn{2}{c}{CASPT2\textsuperscript{b}} &
    \multicolumn{1}{c}{LR-CCSDR(3)\textsuperscript{c}} &
    \multicolumn{2}{c}{LR-CC3\textsuperscript{d}} &
    \multicolumn{2}{c}{Experiment\textsuperscript{e}} \\
    \cmidrule(r){2-3}
    \cmidrule(r){4-5}
    \cmidrule(r){6-6}
    \cmidrule(r){7-8}
    \cmidrule(r){9-10}
    {} &
    {$\Delta E_n$ (eV)} & {$f_n$} &
    {$\Delta E_n$ (eV)} & {$f_n$} &
    {$\Delta E_n$ (eV)} &
    {$\Delta E_n$ (eV)} & {$f_n$} &
    {$\Delta E_n$ (eV)} & {$f_n$} \\
    \midrule

    $1^1\!B_\mathrm{2u}$ ($\pi \to \pi^\ast$) &
      4.66 & 0.070 &
      4.56 & 0.050 &
      5.01 &
      5.03 & 0.085 &
      4.70 & \multicolumn{1}{c}{0.109} \\

    $2^1\!B_\mathrm{3u}$ ($\pi \to \pi^\ast$) &
      6.06 & 1.312 &
      5.54 & 1.337 &
      6.27 &
      6.33 & 1.325 &
      5.89 & \multicolumn{1}{c}{1.3} \\

    $2^1\!B_\mathrm{2u}$ ($\pi \to \pi^\ast$) &
      6.33 & 0.263 &
      5.93 & 0.313 &
      6.51 &
      6.57 & 0.239 &
      6.00 & \multicolumn{1}{c}{---} \\

    \bottomrule
  \end{tabular}

  \vspace{4pt}
  \begin{tablenotes}
    \raggedright
    \footnotesize
    \item \textsuperscript{a} This work.
    \item \textsuperscript{b} Rubio \textit{et al.}\cite{Rubio_1994}: [10e,10o]-CASPT2/ANO-S + single-center Rydberg $(2s,2p,2d)$ at the center of mass.
    \item \textsuperscript{c} Falden \textit{et al.}\cite{Falden_2009}: LR-CCSDR(3)/ANO-S + single-center Rydberg $(2s,2p,2d)$ at the center of mass.
    \item \textsuperscript{d} K\'ann\'ar \textit{et al.}\cite{Kannar_2014}: LR-CC3/TZVP.
    \item \textsuperscript{e} Huebner \textit{et al.}\cite{Huebner_1972}: gas-phase EELS.
  \end{tablenotes}
  \end{threeparttable}
\end{table*}

\subsubsection{Fluorescence cross section and total excitation cross section}

The gas-phase apparent fluorescence cross section\cite{McConkey_1992},
which can serve as an emission-based proxy for the total excitation
cross section when dipole-allowed transitions dominate\cite{Kazakov_2000},
provides an experimental reference for validating the onset and principal
maximum of the calculated cross section. Its onset at $\sim$4.7~eV
corresponds to the threshold of the $1\,^{1}\!B_\mathrm{2u}$
($\pi \to \pi^{\ast}$) state.
As shown in Table~\ref{tab:rep_points}, for a typical
electronic excitation with $r \sim 1$--$10$, the cross-section
maximum for a dipole-allowed transition is expected to appear
at an incident energy of approximately
$1.5$--$1.6\,\Delta E$.
On this basis, the main peak of the fluorescence
cross section observed at $\sim$7.8~eV is assigned to the
$2\,^{1}\!B_\mathrm{3u}$ ($\pi \to \pi^{\ast}$) state, while the
broad shoulder extending over the 9--13~eV range is tentatively
assigned, within the present vertical (Franck--Condon),
dipole-allowed framework, primarily to the $2\,^{1}\!B_\mathrm{2u}$
($\pi \to \pi^{\ast}$) state, with possible contributions from other
dipole-allowed states listed in Table~S1 of the Supplementary
Information.

\begin{figure}[t]
  \centering
  \includegraphics[width=\columnwidth]{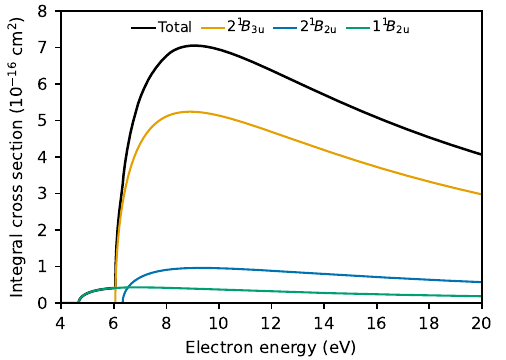}
  \caption{
  The MC-BE/TMMM calculated electron-impact excitation cross sections of naphthalene
  using TD-$\omega$B97X-D target states as a function of incident
  electron energy $T$. Black line: total cross section; orange line:
  $2 ^1B_\text{3u}$ state; blue line: $2 ^1B_\text{2u}$ state; green
  line: $1 ^1B_\text{2u}$ state.}
  \label{fig:naphthalene_state_resolved_cross_section}
\end{figure}

The calculated total excitation cross section, broadened with a
Gaussian function of FWHM $= 0.5$~eV to account for the reported
electron-beam energy spread, is compared with the experimental
fluorescence spectrum\cite{McConkey_1992} in
Fig.~\ref{fig:naphthalene_eels_cross_section}(b). The overall
agreement is satisfactory: the theoretical profile reproduces both the
onset behavior and the position of the principal maximum.
Analysis of
the state-resolved cross sections
(Fig.~\ref{fig:naphthalene_state_resolved_cross_section}) confirms
that the dipole-allowed $2\,^{1}\!B_\mathrm{3u}$ ($\pi \to \pi^{\ast}$)
transition is the dominant state-resolved contributor, accounting for
74\% of the total excitation cross section at
its peak position ($T = 9.07$~eV). This percentage should be viewed as
a calculation-specific state-resolved indicator; the total excitation
cross section is the more relevant quantity for comparison with
experiment. The threshold of the
$1\,^{1}\!B_\mathrm{2u}$ ($\pi \to \pi^{\ast}$) state is located at
$T = 4.66$~eV, in good agreement with the observed fluorescence
onset. The calculated peak of the $2\,^{1}\!B_\mathrm{3u}$
($\pi \to \pi^{\ast}$) cross section appears at $T = 8.90$~eV,
overestimating the experimental fluorescence maximum by
$\sim$1.1~eV. This deviation may correspond to an overestimation of the
excitation energy by $\sim$0.7~eV, which lies within the combined
uncertainties of the TD-$\omega$B97X-D calculation
($\lesssim 0.3$~eV) and the vertical approximation. The
$2\,^{1}\!B_\mathrm{2u}$ ($\pi \to \pi^{\ast}$) state exhibits a broad
plateau in its calculated cross section over the 9--13~eV range,
supporting this tentative assignment.

The remaining discrepancies between the calculated and experimental
profiles can be attributed to two principal sources:
(i) the intrinsic accuracy of the LR-TDDFT excitation energies
($\lesssim 0.3$~eV,
Table~\ref{tab:naphthalene_tddft_eomcc3_exp}) and
(ii) the neglect of vibronic structure within the vertical
approximation\cite{Atkins_2011,Santoro_2008,Barone_2009,Gozem_2022}.
Extension of the present MC-BE framework to incorporate vibronic
effects constitutes a natural direction for future work.

\section{Conclusions}

We have developed a multiconfigurational binary-encounter (MC-BE)
framework that enables direct analytical evaluation of electron-impact
electronic-excitation cross sections from \textit{ab initio}
excited-state data. In the present formulation, the threshold-modified Mott--Massey
(TMMM) approximation provides the Born-type baseline, and the BE
scaling mitigates the near-threshold overestimation characteristic of
the unscaled TMMM model. The key MC-BE extension is the
multiconfigurational evaluation of the effective binding energy
$\langle B_n\rangle$ from LR-TDDFT amplitudes, which enables
excited-state data to be used directly in the analytical scaling
formula without system-specific fitting.

Benchmark calculations for benzene and naphthalene validate the
approach against experimental data. 
For benzene, the MC-BE/TMMM cross
sections for the dominant $1\,{}^{1}\!E_{\mathrm{1u}}$
($\pi\!\to\!\pi^{\ast}$) band reproduce the experimental profile over
the 10--20\,eV range and, for this benchmark, show better agreement
with experiment than the SMC/TCIS results of Ref.~\cite{Falkowski_2023}.
This favorable comparison is benchmark-specific and should not be
interpreted as a general replacement for multichannel scattering
calculations, since MC-BE/TMMM represents scattering effects through
BE-type analytical scaling rather than by explicitly solving the
coupled-channel scattering problem. State-resolved analysis confirms
that the $1\,{}^{1}\!E_{\mathrm{1u}}$ transition provides the dominant
contribution to the total B1-band excitation cross section. 
For naphthalene, the gas-phase apparent fluorescence cross section is
used here as an emission-based proxy under dipole-dominated conditions.
On this basis, the calculated total excitation cross section reproduces
both the onset and the principal maximum without empirical energy
shifts or intensity scaling. State-resolved analysis identifies the
$2\,{}^{1}\!B_\mathrm{3u}$ ($\pi\!\to\!\pi^{\ast}$) state as the dominant
contributor in the present TD-$\omega$B97X-D calculation and supports a
tentative assignment of the broad 9--13\,eV shoulder primarily to the
$2\,{}^{1}\!B_\mathrm{2u}$ ($\pi\!\to\!\pi^{\ast}$) state. The residual
discrepancy of $\sim$1.1\,eV in the peak position may be attributable
to the combined uncertainties of the TD-$\omega$B97X-D excitation
energies ($\lesssim$0.3\,eV) and the vertical (Franck--Condon)
approximation.

Analytic expressions for the peak position and peak height of the
BE/BE$f$-scaled TMMM cross section, parameterized by the
dimensionless binding ratio $r = \langle B_n \rangle / \Delta E_n$,
provide useful diagnostic relations: for a typical electronic excitation ($r \sim 1$--$10$), the
cross-section maximum appears at
$T \approx 1.5$--$1.6\,\Delta E_n$ with a peak height of
approximately $0.08$--$0.28\,\sigma_{\mathrm{TMMM}}^{\ast}$. These relations offer a
practical means of estimating excitation energies from experimental
cross-section data.

The methodology is not restricted to LR-TDDFT. Because the effective
binding energy $\langle B_n \rangle$ is defined through
amplitude-weighted occupied-orbital contributions, 
the framework admits a formal extension to other
wave-function-based excited-state theories, provided that compatible
amplitudes and well-defined orbital energies are available for the
chosen reference function. For the benzene
$1\,{}^{1}\!E_{\mathrm{1u}}$ benchmark, the MC-BE/TMMM result is weakly
sensitive to the choice between the two long-range corrected
functionals examined here, supporting the robustness of the total
cross section for this test case.

Two principal directions for future work emerge from the present
study. First, incorporation of Franck--Condon factors and
Herzberg--Teller vibronic
coupling\cite{Santoro_2008,Barone_2009,Gozem_2022} would account for
vibronic effects beyond the present vertical treatment and improve
band-shape predictions, while spin--orbit
coupling\cite{Marian_2012,Marian_2021,Kotaru_2023,Kamiya_2018} would
allow transitions that acquire finite oscillator strength through
spin--orbit mixing to be treated, particularly for transitions such
as the $1\,{}^{1}\!B_\mathrm{2u}$ state of naphthalene, where the
Franck--Condon envelope significantly broadens the spectral
profile\cite{Atkins_2011}.
Second, systematic application to larger polycyclic aromatic
hydrocarbons and heteroaromatic
systems\cite{Kazakov_2000} would further establish the scope and
predictive capability of the method across diverse molecular classes.

\begin{acknowledgments}
This work was supported by the Institute for Quantum Chemical
Exploration grant-in-aid (No.~R07Josei003).
The authors acknowledge the use of computational resources provided by Research Center for Computational Science, Okazaki, Japan (Projects: 25-IMS-C099 and 26-IMS-C092). 
The authors are grateful to Drs. ~Tetsuya Fukushima, Hisao Nakamura, and Ryosuke Senga at AIST, Japan for valuable discussions and helpful comments.

\end{acknowledgments}

\appendix

\section{Generalization to CI-type wave functions}
\label{app:CI_Bn}

For a general CI-type excited-state wave function that
includes single, double, and higher excitations,
\begin{equation}
\ket{\Psi_n} = \sum_{ia} X_{ia}^{n} \ket{\Phi_i^a}
             + \sum_{\substack{i<j \\ a<b}}
               X_{ijab}^{n} \ket{\Phi_{ij}^{ab}}
             + \cdots,
\end{equation}
where $i,j,\ldots$ run over occupied and $a,b,\ldots$ over
virtual orbitals of the reference function, the normalization
constant is defined as
\begin{equation}
C_n = \sum_{i}^{\mathrm{occ}}\sum_{a}^{\mathrm{vir}}
      \left|X_{ia}^{n}\right|^2
    + \sum_{\substack{i<j}}^{\mathrm{occ}}
      \sum_{\substack{a<b}}^{\mathrm{vir}}
      \left|X_{ijab}^{n}\right|^2
    + \cdots.
\label{eq:Cn_CI}
\end{equation}
The effective binding energy is then defined
as\cite{Klinkusch_2009,Coccia_2017}:
\begin{align}
\langle B_n\rangle_{\mathrm{CI}}
= -\frac{1}{C_n}
  \biggl[
  &\sum_{i}^{\mathrm{occ}}\sum_{a}^{\mathrm{vir}}
    \left|X_{ia}^{n}\right|^2 \epsilon_i
  \notag\\
  &+ \sum_{\substack{i<j}}^{\mathrm{occ}}
    \sum_{\substack{a<b}}^{\mathrm{vir}}
    \left|X_{ijab}^{n}\right|^2
    (\epsilon_i + \epsilon_j)
  + \cdots
  \biggr].
\label{eq:Bn_CI_general}
\end{align}

When only singly excited configurations are retained,
Eqs.~\eqref{eq:Cn_CI} and \eqref{eq:Bn_CI_general} reduce
to
\begin{equation}
\langle B_n\rangle_{\mathrm{CIS}}
= -\frac{1}{C_n}
  \sum_{i}^{\mathrm{occ}}\sum_{a}^{\mathrm{vir}}
  \left|X_{ia}^{n}\right|^2 \epsilon_i
\label{eq:Bn_CIS}
\end{equation}
which is the CI-singles (CIS) analogue of the LR-TDDFT
expression Eq.~\eqref{eq:Bn_LR_TDDFT}. The two become
identical under the Tamm--Dancoff approximation
(TDA)\cite{Tsuneda_2014}, in which the de-excitation
amplitudes vanish ($Y_{ia}^n = 0$) and the LR-TDDFT
amplitudes $X_{ia}^n$ coincide with the CIS coefficients\cite{Wozniak_2024}.
Equation~\eqref{eq:Bn_CIS} is therefore directly applicable
to DFT-based single-excitation calculations performed within
the TDA framework.

Extensions to excited-state theories such as EOM-CC, LR-CC,
SAC-CI, and ADC\cite{Jensen_2017} follow the same
principle, provided that a consistent set of amplitudes and
occupied-orbital energies is available for the chosen
reference function.

\section{Analytic Peak Positions and Peak Height of TMMM}
\label{sec:tmmm_peak}

\subsection{TMMM: Peak position and peak height}

The TMMM integral cross section for the transition to excited state $n$
is
\begin{align}
\sigma_n^{\mathrm{TMMM}}(T)
&= \Theta(T-\Delta E_n)\,
  \frac{2\pi\,g_n f_n}{\Delta E_n\,T}\,
  \ln\!\left(
    \frac{\sqrt{\Delta E_n}}{\sqrt{T}-\sqrt{T-\Delta E_n}}
  \right), \notag\\
&\qquad (T>\Delta E_n).
\label{eq:tmmm_sigma}
\end{align}

To locate the peak, differentiate $\sigma_n^{\mathrm{TMMM}}(T)$ with
respect to $T$ and set the derivative to zero. Introducing the
dimensionless energy $s\equiv T/\Delta E_n$ and the auxiliary
variable
\begin{equation}
\chi \equiv \sqrt{\frac{T}{T-\Delta E_n}}
= \sqrt{\frac{s}{s-1}},\qquad \chi>1,
\label{eq:tmmm_chi_def}
\end{equation}
the stationarity condition
$\mathrm{d}\sigma_n^{\mathrm{TMMM}}/\mathrm{d}T=0$ reduces to the
$T$-independent transcendental equation
\begin{equation}
\chi = \ln\!\left(\frac{\chi+1}{\chi-1}\right),\qquad \chi>1,
\label{eq:tmmm_chi_eq}
\end{equation}
whose unique solution is $\chi\simeq 1.5434$. Hence the peak position
is
\begin{equation}
s^{\ast}_{\mathrm{TMMM}}
=\frac{T^{\ast}_{\mathrm{TMMM}}}{\Delta E_n}
=\frac{\chi^2}{\chi^2-1}
\simeq 1.7235,
\label{eq:tmmm_sstar}
\end{equation}
i.e., $T^{\ast}_{\mathrm{TMMM}} =
s^{\ast}_{\mathrm{TMMM}}\,\Delta E_n$.

At this energy, the peak height becomes
\begin{align}
\sigma_n^{\mathrm{TMMM}}
   \!\left(T^{\ast}_{\mathrm{TMMM}}\right)
&= \frac{2\pi}{s^{\ast}_{\mathrm{TMMM}}}\,
   \frac{\ln\!\big(\sqrt{\chi^2-1}/(\chi-1)\big)}{(\Delta E_n)^2}\,
   g_n f_n
\notag\\
&\equiv \frac{\mathcal{S}_{\mathrm{TMMM}}}{(\Delta E_n)^2}\,g_n f_n,
\qquad \mathcal{S}_{\mathrm{TMMM}}\approx 2.8129.
\label{eq:tmmm_peak_height}
\end{align}

\section{BE-scaled TMMM: Dimensionless Formulation, Peak Analysis, and Ratios}
\label{app:dim_peak}

\subsection{Dimensionless formulation and the peak position}
\label{subsec:nondim_app}

Introduce
\begin{equation}
s \equiv \frac{T}{\Delta E_n},\qquad
r \equiv \frac{\langle B_n\rangle}{\Delta E_n},\qquad s\ge 1,
\end{equation}
and use
$\sqrt{s}-\sqrt{s-1}=\bigl(\sqrt{s}+\sqrt{s-1}\bigr)^{-1}$ to define
\begin{equation}
\beta(s) \equiv \sqrt{s}+\sqrt{s-1},
\qquad
\ln\!\left(\frac{1}{\sqrt{s}-\sqrt{s-1}}\right)=\ln \beta(s).
\end{equation}

Up to transition-specific constants that do not affect the peak
location, define the objective function
\begin{equation}
\mathcal{J}(s;r)\equiv
\ln\!\bigl[\ln\beta(s)\bigr]-\ln s-\ln(1+s+r).
\end{equation}
The unique peak $s^{\ast}(r)=T^{\ast}/\Delta E_n$ follows from the
stationarity condition
\begin{equation}
\frac{d}{ds}\,\mathcal{J}(s;r)=0,
\label{eq:stat_raw}
\end{equation}
which expands to
\begin{equation}
\frac{\beta'(s)}{\beta(s)\,\ln \beta(s)}
=\frac{1}{s}+\frac{1}{1+s+r}
=\frac{1+r+2s}{s(1+s+r)},
\label{eq:sstar_implicit}
\end{equation}
with
\[
\beta'(s)=\frac{1}{2\sqrt{s}}+\frac{1}{2\sqrt{s-1}}.
\]
Since $\eta_n$ is independent of $T$, it cancels from the stationarity
condition. Therefore, BE and BE$f$ share the same peak position
$s^{\ast}(r)$ and $T^{\ast}$ for a fixed electronic-structure input.

\subsection{Peak attenuation relative to TMMM}
\label{subsec:attenuation_app}

A natural measure of the BE attenuation at the \emph{same} energy
$T^{\ast}$ is
\begin{equation}
R_{\mathrm{BE}}(r)
\equiv
\frac{\sigma^{\mathrm{BE}}(T^{\ast})}{\sigma^{\mathrm{TMMM}}(T^{\ast})}
=\frac{1}{1+r+s^{\ast}(r)}.
\label{eq:R_sameT}
\end{equation}
A complementary peak-height ratio compares the global maxima:
\begin{equation}
\mathcal{R}_{\mathrm{peak}}(r)
\equiv
\frac{\max_T \sigma^{\mathrm{BE}}(T)}{\max_T \sigma^{\mathrm{TMMM}}(T)}
=
\frac{ \displaystyle \frac{\ln \beta\!\bigl(s^{\ast}(r)\bigr)}{s^{\ast}(r)\,[1+r+s^{\ast}(r)]} }
     { \displaystyle \frac{\ln \beta(s_\infty)}{s_\infty} }.
\label{eq:R_peak}
\end{equation}

\subsection{Asymptotics for \texorpdfstring{$s^{\ast}(r)$}{s*(r)} and \texorpdfstring{$R$}{R}}
\label{subsec:asymptotics_app}

\begin{figure}[htbp]
  \centering
  \includegraphics[width=\columnwidth]{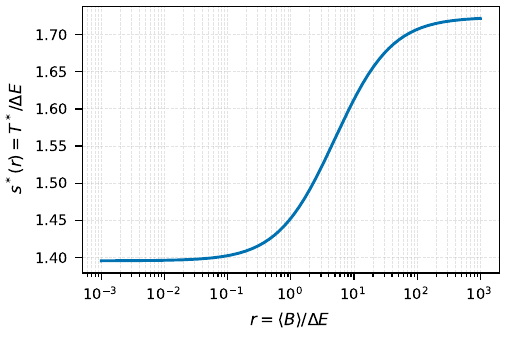}
  \caption{Peak position multiplier $s^*(r)=T^*/\Delta E$ as a
  function of $r=\langle B\rangle/\Delta E$ for the BE/BE$f$-scaled
  TMMM model. The curve increases monotonically from
  $s^*(0)=1.3955$ toward the unscaled TMMM limit
  $s_\infty=1.7235$ as $r\to\infty$.}
  \label{fig:sstar_vs_r}
\end{figure}

Numerical solution of Eq.~\eqref{eq:sstar_implicit} yields
\begin{align}
s^{\ast}(0) &= 1.395531, \notag\\
s_\infty &\equiv \lim_{r\to\infty}s^{\ast}(r) = 1.723536.
\label{eq:s_limits}
\end{align}
The resulting monotonic dependence of $s^\ast(r)$ on $r$ is shown in
Fig.~\ref{fig:sstar_vs_r}.
A first-order small-$r$ expansion follows from implicit
differentiation of Eq.~\eqref{eq:stat_raw}:
\begin{equation}
s^{\ast}(r)=s^{\ast}(0)+c\,r+\mathcal{O}(r^2),\qquad c=0.07147.
\label{eq:s_smallr}
\end{equation}
At large $r$ we obtain the matched-asymptotic form
\begin{equation}
s^{\ast}(r)=s_\infty-\frac{\kappa}{r}
+\mathcal{O}\!\bigl(r^{-2}\bigr),
\qquad \kappa\simeq 1.7448.
\label{eq:s_larger}
\end{equation}
Inserting Eqs.~\eqref{eq:s_smallr}--\eqref{eq:s_larger} into
Eq.~\eqref{eq:R_sameT} gives, respectively,
\begin{align}
R_{\mathrm{BE}}(r)
&= \frac{1}{1+s^{\ast}(0)}
   -\frac{1+c}{\bigl[1+s^{\ast}(0)\bigr]^2}\,r
   +\mathcal{O}(r^2),
\label{eq:R_smallr}\\
R_{\mathrm{BE}}(r)
&= \frac{1}{1+r+s_\infty} + \mathcal{O}\!\bigl(r^{-3}\bigr)
 \sim \frac{1}{r}-\frac{1+s_\infty}{r^2}+\cdots.
\label{eq:R_larger}
\end{align}
Analogous substitutions in Eq.~\eqref{eq:R_peak} provide compact
asymptotics for $\mathcal{R}_{\mathrm{peak}}(r)$.

\subsection{Numerical evaluation}
\label{app:numerics}

\begin{figure}[htbp]
  \centering
  \includegraphics[width=\columnwidth]{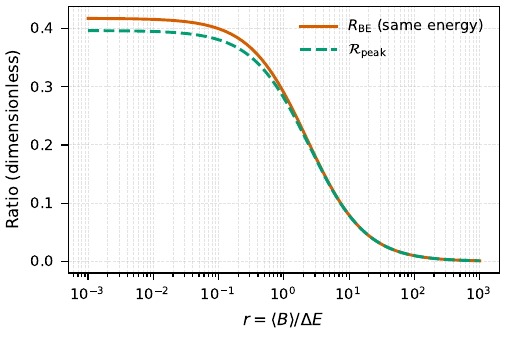}
  \caption{Comparison between the same-energy ratio
    $R_{\mathrm{BE}}(r)$ and the peak-height ratio
    $\mathcal{R}_{\mathrm{peak}}(r)$. Both share the
    $(1+r+s_\infty)^{-1}$ decay at large $r$, while they differ
    quantitatively in the small-to-intermediate $r$ regime.}
  \label{fig:R_both}
\end{figure}

Define
\begin{equation}
F(s;r)\equiv
\frac{\beta'(s)}{\beta(s)\,\ln \beta(s)}
-\frac{1}{s}-\frac{1}{1+s+r}.
\end{equation}
For any $r\ge 0$, $F(\cdot\,;r)$ is continuous on $s\ge 1$ and has a
unique zero in $s\in[1.05,10]$; a bracketing solver, such as Brent's
method, converges rapidly to $s^{\ast}(r)$. Substituting
$s^{\ast}(r)$ into Eqs.~\eqref{eq:R_sameT} and
\eqref{eq:R_peak} gives smooth curves for the attenuation ratios,
as shown in Fig.~\ref{fig:R_both}.

\subsubsection{Physical interpretation and state-resolved application}
\label{subsec:interpretation}

The BE/BE$f$ prefactor
$\Delta E_n/(\Delta E_n+\langle B_n\rangle+T)$ mitigates the
near-threshold overestimation of the TMMM cross section. For finite
binding ratios $r=\langle B_n\rangle/\Delta E_n$, this scaling shifts
the peak to a lower dimensionless energy than the unscaled TMMM value
$s^{\ast}_{\mathrm{TMMM}}\simeq 1.7235$.
The peak position varies only
moderately, from $s^\ast(0)\approx 1.3955$ to
$s_\infty\approx 1.7235$, as $r$ increases from $0$ to $\infty$.

Because BE$f$ differs from BE only through the state-specific
multiplicative factor $\eta_n$, the BE and BE$f$ scalings give the
same peak position for a given transition, while the peak amplitude is
scaled by $\eta_n$. For a set of contributing excited states, one
evaluates $r_n=\langle B_n\rangle/\Delta E_n$ for each state, solves
Eq.~\eqref{eq:sstar_implicit} for $s_n^\ast(r_n)$, and obtains the
state-resolved peak energies
$T_n^\ast=\Delta E_n s_n^\ast(r_n)$. The total excitation profile is
then constructed by summing the state-resolved cross sections. Thus,
the BE and BE$f$ choices affect the relative amplitudes through
$\eta_n$, but not the peak-position condition for a fixed set of
electronic-structure inputs.

\section{Comparison between theoretical and experimental IPs}
\label{app:ip_dft_vs_exp}

\begin{table}[htbp]
\centering
\caption{Comparison of ionization potentials from the orbital energy
of HOMO, $\Delta$SCF, and experiment.}
\label{tab:ionization_potentials}
\begin{tabular}{
    l
    S[table-format=2.2]
    S[table-format=2.2]
    S[table-format=2.2]
}
\toprule
Molecule & {$-\epsilon_{\text{HOMO}}$ (eV)} & {$\Delta$SCF (eV)} & {Experiment (eV)} \\
\midrule
Benzene      & 9.07 & 9.27 & 9.24\textsuperscript{a} \\
Naphthalene  & 7.98 & 8.04 & 8.00\textsuperscript{b} \\
\bottomrule
\end{tabular}

\vspace{4pt}
\footnotesize
\textsuperscript{a} Åsbrink \textit{et al.}\cite{Asbrink_1970};
\textsuperscript{b} Yamauchi \textit{et al.}\cite{Yamauchi_1998}.
\end{table}

\section{Dependence of the MC-BE/TMMM cross section on the choice of exchange--correlation functional}
\label{app:xc_dependence}

To assess the sensitivity of the MC-BE/TMMM cross sections to
the choice of exchange--correlation functional for the benzene
B1-band benchmark, we recomputed the benzene cross sections with a
second long-range corrected hybrid functional, LC-$\omega$HPBE
\cite{Henderson_2009}, using the same basis set and computational
protocol as described in Sec.~\ref{sec:comp_details}.

\begin{figure}[tbp]
  \centering
  \includegraphics[width=\columnwidth]{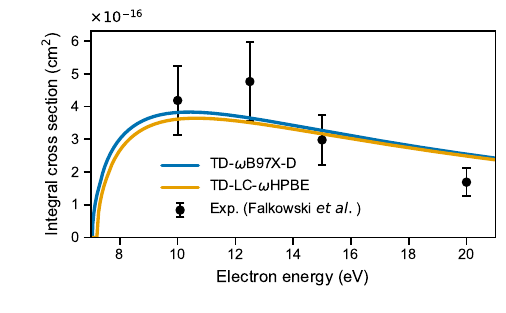}
  \caption{(Color online)
  Integral cross section for electron-impact excitation in the B1
  band of benzene for $T = 7$--21~eV. Experimental data (black
  circles, Ref.~\cite{Falkowski_2023}) are compared with the
  present MC-BE/TMMM results using TD-$\omega$B97X-D (blue) and
  TD-LC-$\omega$HPBE (orange) target states.}
  \label{fig:xc_functional_dependence}
\end{figure}

Figure~\ref{fig:xc_functional_dependence} compares the
integral cross sections in the B1 band of benzene obtained
with $\omega$B97X-D\cite{Chai_2007} and
LC-$\omega$HPBE\cite{Henderson_2009}. The two profiles are
nearly indistinguishable over the entire 7--21~eV range,
confirming the qualitative robustness of the MC-BE/TMMM
results. The corresponding state-resolved and total peak
properties are summarized in
Table~\ref{tab:benzene_peaks_xc_dependence}: the total peak
cross sections agree to within $\sim$5\,\%
($3.83 \times 10^{-16}$~cm$^2$ for $\omega$B97X-D versus
$3.64 \times 10^{-16}$~cm$^2$ for LC-$\omega$HPBE), and the
peak positions agree to within 0.25~eV. Although the
individual oscillator strengths of the
$1\,{}^{1}\!E_{\mathrm{1u}}$ and $2\,{}^{1}\!E_{\mathrm{1u}}$ states differ at
the $\sim$15\,\% level between the two functionals---a
consequence of differing valence--Rydberg configuration
mixing in this quasi-degenerate
pair\cite{Lorentzon_1995,Nakatsuji_2026}---their partial sum,
which is the quantity that enters Eq.~\eqref{eq:sigma_sum},
is conserved to within $\sim$2\,\% (1.23 versus 1.25).
This robustness can be attributed to the fact that both
long-range corrected functionals yield similarly accurate
orbital energies and, consequently, similar effective binding
energies $\langle B_n \rangle$
(Table~\ref{tab:benzene_peaks_xc_dependence}) entering the
BE/BE$f$ scaling. 

\begin{table}[htbp]
\centering
\caption{State-resolved and total peak properties of the
MC-BE/TMMM integral cross sections of benzene calculated with
two long-range corrected hybrid functionals,
$\omega$B97X-D\cite{Chai_2007} and
LC-$\omega$HPBE\cite{Henderson_2009}.
 Reported quantities are
the vertical excitation energy $\Delta E_n$, oscillator
strength $f_n$, effective binding energy
$\langle B_n \rangle$ [Eq.~\eqref{eq:Bn_LR_TDDFT}], peak
position $T^\ast_n$, and peak cross-section value
$\sigma^\ast_n(T^\ast_n)$.}
\label{tab:benzene_peaks_xc_dependence}
\begin{tabular}{lccccc}
\toprule
State & $\Delta E_n$ (eV) & $f_n$ & $\langle B_n \rangle$ (eV) & $T^\ast_n$ (eV) & $\sigma^\ast_n(T^\ast_n)$ (cm$^2$) \\
\midrule
\multicolumn{6}{l}{\textit{TD-$\omega$B97X-D}} \\
$1^1\!A_\mathrm{2u}$ ($\pi \to 3p_\sigma$) & 7.09 & 0.056 & 9.16 & 10.39 & $1.67 \times 10^{-17}$ \\
$1^1\!E_\mathrm{1u}$ ($\pi \to \pi^\ast$)  & 7.05 & 1.085 & 9.46 & 10.34 & $3.26 \times 10^{-16}$ \\
$2^1\!E_\mathrm{1u}$ ($\pi \to 3p_\pi$)    & 7.35 & 0.145 & 9.23 & 10.76 & $4.08 \times 10^{-17}$ \\
Total & --- & --- & --- & 10.39 & $3.83 \times 10^{-16}$ \\
\midrule
\multicolumn{6}{l}{\textit{TD-LC-$\omega$HPBE}} \\
$1^1\!A_\mathrm{2u}$ ($\pi \to 3p_\sigma$) & 7.60 & 0.062 & 9.98  & 11.15 & $1.62 \times 10^{-17}$ \\
$1^1\!E_\mathrm{1u}$ ($\pi \to \pi^\ast$)  & 7.21 & 1.240 & 10.40 & 10.61 & $3.46 \times 10^{-16}$ \\
$2^1\!E_\mathrm{1u}$ ($\pi \to 3p_\pi$)    & 7.87 & 0.008 & 9.92  & 11.53 & $1.87 \times 10^{-18}$ \\
Total & --- & --- & --- & 10.64 & $3.64 \times 10^{-16}$ \\
\bottomrule
\end{tabular}
\end{table}

\bibliography{refs_260629}

\section*{Data Availability}
The data that support the findings of this study are available from
the corresponding author upon reasonable request.

\end{document}